\documentclass[aps,prb,showpacs,groupedaddress]{revtex4}
\usepackage{amsmath,graphicx}
\usepackage{epsfig}
\usepackage{bm}
\usepackage[next]{inputenc}

\begin{document}
\title{Comment on "Dielectric screening and plasmons in AA-stacked bilayer graphene"}
\author{Yawar Mohammadi}
\email[]{yawar.mohammadi@gmail.com} \affiliation{Young Researchers
and Elite Club, Kermanshah Branch, Islamic Azad University,
Kermanshah, Iran }



\date{\today}
\pacs{71.45.Lr, 73.22.Pr, 71.51.Gm} \maketitle

In reference [1], authors consider dielectric screening and
plasmons in AA-stacked bilayer graphene. Some equations and
results obtained in this paper seem to be incorrect. In this
comment, we demonstrate this claim and present our results. At
first we summarize our comments. The obtained eigenfunctions for
the unbiased case, Eqs. (7) and (8), don't satisfy the
Schr\"{o}dinger equation. Equations (31), (32) and (33) don't
satisfy the Schr\"{o}dinger equation and also don't reduce to the
corresponding results in the zero limit of V (the applied
perpendicular electric potential). The overlap of the electron and
hole wavefunctions, Eq. (35), can not be concluded from Eq. (33)
and also it doesn't reduce to Eq. (17) in the zero limit of V,
while it must be so. Furthermore, we show that this result for the
overlap of the electron and hole wavefunctions is not correct. The
main comment is that the low energy limit of the optical and
acoustic plasmon dispersion relations, Eqs. (36), (37) and (39),
can not be obtained by making use of Eq. (35). Plots of figure 4
can not be correct, since they have been obtained by making use of
Eq. (35). Some of these mistakes arise from the used unitary
transformation. For example, even if the calculations are done
correctly, the biased-case eigenfunctions don't reduce to the
corresponding result in the zero limit of V. In the rest of this
comment we present our results.

Starting from Eq. (1) of Ref. [1] for $\mathcal{H}$, first we
obtain a unitary transformation matrix, which can be used to
transform this Hamiltonian matrix into a block-diagonalized form
similar to Eq. (5) of Ref. [1] denoted, in this comment, by
$\mathcal{H}_{bd}$. Let us suppose that $X^{\dag}\mathcal{H}X$ and
$Y^{\dag}\mathcal{H}_{bd}Y$ are the transformations that
diagonalize $\mathcal{H}$ and $\mathcal{H}_{bd}$ respectively.
 These matrixes, $X$ and $Y$, can be written as
\begin{eqnarray}
X=(
\begin{array}{cccc}
\Phi_{+,-}(\mathbf{k}) & \Phi_{-,-}(\mathbf{k}) &
\Phi_{+,+}(\mathbf{k}) & \Phi_{-,+}(\mathbf{k})
\end{array}), \\ \nonumber
Y=(
\begin{array}{cccc}
\Psi_{+,-}(\mathbf{k}) & \Psi_{-,-}(\mathbf{k}) &
\Psi_{+,+}(\mathbf{k}) & \Psi_{-,+}(\mathbf{k})
\end{array}),
\label{eq:01}
\end{eqnarray}
where
\begin{eqnarray} \Phi_{\lambda,s}(\mathbf{k})= \frac{1}{2}\left(
\begin{array}{c}
1  \\
\lambda e^{i\phi_{\mathbf{k}}}  \\
-s   \\
-s\lambda e^{i\phi_{\mathbf{k}}}  \\
\end{array}
\right),\label{eq:02}
\end{eqnarray}
are eigenfunctions of $\mathcal{H}$, Eq. (1) of Ref. [1], and
\begin{eqnarray} \Psi_{\lambda,-}(\mathbf{k})= \frac{1}{\sqrt{2}}\left(
\begin{array}{c}
1  \\
\lambda e^{i\phi_{\mathbf{k}}}  \\
0 \\
0 \\
\end{array}
\right), \Psi_{\lambda,+}(\mathbf{k})= \frac{1}{\sqrt{2}}\left(
\begin{array}{c}
0 \\
0 \\
1 \\
\lambda e^{i\phi_{\mathbf{k}}}  \\
\end{array}
\right),\label{eq:03}
\end{eqnarray}
are eigenfunctions of $\mathcal{H}_{bd}$, Eq. (5) of Ref. [1],
with $s=\pm$ and $\lambda=\pm$. It is easy to show that these
eigenfunctions satisfy the corresponding Schr\"{o}dinger equations
with $\varepsilon_{\lambda,s}(\mathbf{k})=st_{1}+\lambda v_{F}k $.
Since $X^{\dag}\mathcal{H}X$ and $Y^{\dag}\mathcal{H}_{bd}Y$ are
equal, we can conclude that
$YX^{\dag}\mathcal{H}XY^{\dag}=\mathcal{H}_{bd}$. So we reach a
unitary transformation ($U^{-1}\mathcal{H}U=\mathcal{H}_{bd}$)
which transforms $\mathcal{H}$ into a block-diagonalized form as
Eq. (5) of Ref. [1]. Here, $U$ is given by
\begin{eqnarray}
U= \frac{1}{\sqrt{2}}\left(
\begin{array}{cccccc}
1 & 0 & 1  & 0  \\
0 & 1 & 0  & 1  \\
1 & 0 & -1 & 0  \\
0 & 1 & 0  & -1 \\
\end{array}
\right).\label{eq:04}
\end{eqnarray}
In accordance to this unitary transformation, we do
$\mathbf{V}(q)=\mathcal{U}^{-1}\mathbf{\tilde{V}}(q)\mathcal{U}$
to transform the Coulomb interaction matrix from the layer1/layer2
basis to the bonding/antibonding basis, where
\begin{eqnarray}
\mathcal{U}= \left(
\begin{array}{cc}
1 & 1 \\
1 & -1 \\
\end{array}
\right).\label{eq:05}
\end{eqnarray}
Notice that this unitary transformation leads to the same results
as Eqs. (14)-(28) of Ref. [1].

Now we calculate the eigenfunctions of the Hamiltonian introduced
in Eq. (29) of Ref. [1], subjected to our unitary transformation.
To obtain these eigenfunctions, first we obtain the eigenfunctions
of $H_{V}$ denoted here by $\psi_{\lambda,s}^{V}(\mathbf{k})$ and
then calculate
$\Psi_{\lambda,s}^{V}(\mathbf{k})=U^{-1}\psi_{\lambda,s}^{V}(\mathbf{k})$
which are the eigenfunctions of $\mathcal{H}^{V}=U^{-1}H^{V}U$. So
we have
\begin{eqnarray} \Psi^{V}_{\lambda,s}(\mathbf{k})= \frac{1}{2\sqrt{2t_{1}^{V}(t_{1}^{V}-sV)}}\left(
\begin{array}{c}
t_{1}+(V-st_{1}^{V}) \\
(t_{1}+(V-st_{1}^{V}))\lambda e^{i\phi_{\mathbf{k}}}  \\
t_{1}-(V-st_{1}^{V}) \\
(t_{1}-(V-st_{1}^{V}))\lambda e^{i\phi_{\mathbf{k}}} \\
\end{array}
\right). \label{eq:06}
\end{eqnarray}
It is easy to show that these eigenfunctions, in zero limit of
$V$, reduces to Eq. (3) of this comment. While if we use the
unitary transformation matrix of Re. [1], the obtained
eigenfunctions, in the zero limit of $V$, do not reproduce the
correct eigenfunction introduced in Eq. (3).

By using Eq. (\ref{eq:06}) and calculating 1-loop
polarization~\cite{b.02}, we obtain
\begin{eqnarray}
\Pi^{V}_{s,s^{'};\lambda,\lambda^{'}}(\mathbf{q})=-\frac{g_{\sigma}g_{\upsilon}}{L^{2}}
\sum_{\mathbf{k}
}\frac{f_{\lambda,s}(\mathbf{k})-f_{\lambda^{'},s^{'}}(\mathbf{k}+\mathbf{q})}
{\omega+\varepsilon^{V}_{\lambda,s}(\mathbf{k})-\varepsilon^{V}_{\lambda^{'},s^{'}}(\mathbf{k}+\mathbf{q})+i\delta}
\mathcal{F}^{V}_{ss^{'};\lambda\lambda^{'}}(\mathbf{k},\mathbf{k}+\mathbf{q}),\label{eq:07}
\end{eqnarray}
where
$\mathcal{F}^{V}_{ss^{'};\lambda\lambda^{'}}({\mathbf{k}\mathbf{k}^{'}})=|\langle
\Psi_{\lambda^{'},s^{'}}^{V}(\mathbf{k}^{'})
|e^{i(\mathbf{k}-\mathbf{k}^{'}).\mathbf{r}}|
\Psi_{\lambda,s}^{V}((\mathbf{k})) \rangle|^{2}$ is the overlap of
the electron and hole wavefunctions which is given by
\begin{eqnarray}
\mathcal{F}^{V}_{ss^{'};\lambda\lambda^{'}}(\mathbf{k},\mathbf{k}+\mathbf{q})=
\frac{1+\lambda\lambda^{'}\cos(\phi_{\mathbf{k}}-\phi_{\mathbf{k}+\mathbf{q}})}{2},\label{eq:08}
\end{eqnarray}
if $s=s^{'}$ and
$\mathcal{F}^{V}_{ss^{'};\lambda\lambda^{'}}(\mathbf{k},\mathbf{k}+\mathbf{q})=0$
when $s\neq s^{'}$. This equation shows that the dynamical
polarization of the biased AA-stacked BLG, is equal to that of
doped single layer graphene with $\mu=t_{1}^{V}$,
$\Pi^{\mu=t^{V}_{1}}_{SLG}(\mathbf{q},\omega)$. By making use of
Eqs. (6) and (7), we obtain\cite{b.03} a closed analytical
expression for the real part of the polarization function
 as
\begin{eqnarray}
\Pi^{V}(\mathbf{q},\omega)=-\frac{g_{\sigma}g_{\upsilon}t^{V}_{1}}{2\pi
v^{2}_{F}
}+\frac{g_{\sigma}g_{\upsilon}q^{2}}{16\pi\sqrt{\omega^{2}-v^{2}_{F}q^{2}}}
[G(\frac{2t^{V}_{1}+\omega}{v_{F}q})-G(\frac{2t^{V}_{1}-\omega}{v_{F}q})]
,\label{eq:09}
\end{eqnarray}
with $G(x)=x\sqrt{x^{2}-1}-\cosh^{-1}(x)$ while
Im$\Pi^{V}(\mathbf{q},\omega)=0$. Notice that each cone
contributes independently and their contributions are equal. This
relation is valid in the region of the spectrum
($v_{F}q<\omega<2t^{V}_{1}-v_{F}q$) where it is claimed that Eq.
(36) and (37) of Ref. [1] are the low energy dispersion relation
for the optical and acoustic plasmon modes. It is easy to show
that Eq. (36) and (37) of Ref. [1] are obtained by making use of
our results, while they can not be concluded from corresponding
relation in Ref. [1], Eq. (35). To demonstrate our claim, we
rewrite Eq. (35) of Ref. [1] as
\begin{eqnarray}
\mathcal{F}^{V}_{ss^{'};\lambda\lambda^{'}}(\mathbf{k},\mathbf{k}+\mathbf{q})&=&\frac{1-\beta^{2}_{s}(t_{1},V)}{2[1+\beta^{2}_{s}(t_{1},V)]}
+\frac{\beta^{2}_{s}(t_{1},V)}{1+\beta^{2}_{s}(t_{1},V)}\frac{1+\lambda\lambda^{'}\cos(\phi_{\mathbf{k}}-\phi_{\mathbf{k}+\mathbf{q}})}{2},\label{eq:10}
\end{eqnarray}
for $s=s^{'}$. So we can obtain an analytical relation for the
real part of the dynamical polarization function as
\begin{eqnarray}
\Pi^{V}(\mathbf{q},\omega)=\frac{\beta^{2}_{s}(t_{1},V)}{1+\beta^{2}_{s}(t_{1},V)}\Pi^{\mu=t^{V}_{1}}_{SLG}(\mathbf{q},\omega)
+\frac{1-\beta^{2}_{s}(t_{1},V)}{[1+\beta^{2}_{s}(t_{1},V)]}\Pi^{'}(\mathbf{q},\omega),\label{eq:11}
\end{eqnarray}
being valid in the region $v_{F}q<\omega<2t^{V}_{1}-v_{F}q$, where
\begin{eqnarray}
\Pi^{'}(\mathbf{q},\omega)=\frac{g_{\sigma}g_{\upsilon}q^{2}}{16\pi\sqrt{\omega^{2}-v^{2}_{F}q^{2}}}
[G(\frac{2t^{V}_{1}+\omega}{v_{F}q})-G(\frac{2t^{V}_{1}-\omega}{v_{F}q})]-\frac{g_{\sigma}g_{\upsilon}q^{2}\sqrt{\omega^{2}-v^{2}_{F}}}
{16\pi}[\cosh^{-1}(\frac{2t^{V}_{1}+\omega}{v_{F}q})-\cosh^{-1}(\frac{2t^{V}_{1}-\omega}{v_{F}q})].\label{eq:12}
\end{eqnarray}
One can show easily that Eqs. (36) and (37) of Ref. [1] can not be
concluded from Eqs. {\ref{eq:11} and {\ref{eq:12} of this comment
obtained from Eq. (35) of Ref. [1].

In summary, we showed that some equations obtained in Ref. [1] are
not correct. Furthermore, we present our results which seems to be
correct.



\section*{References}
\begin{thebibliography}{99}

\bibitem{b.01} R. Roldan, L. Brey, Phys. Rev. B {\bf 88}, 115420 (2013).
\bibitem{b.02} Y. Mohammadi, R. Moradian and F. Sirzadi Tabar, Solid State Commun. {\bf 193}, 1
(2014).
\bibitem{b.03} B. Wunsch, T. Stauber, F. Sols, F. Guinea, New. J. Phys. {\bf
8}, 318 (2006).

\end{thebibliography}
\end{document}